\documentclass[11pt,a4paper]{article}

\usepackage{times} 
\usepackage{graphicx}
\usepackage{caption}
\usepackage{subcaption}
\usepackage{amssymb}
\usepackage{amsthm}
\usepackage{amsmath}

\usepackage{lineno}
\usepackage[margin=2.5cm]{geometry}

\usepackage[utf8]{inputenc} 

\begin{document}

\title{Estimation of Li-ion degradation test sample sizes required to understand cell-to-cell variability}

\date{}

\author{Philipp~Dechent\thanks{Authors contributed equally} \thanks{Electrochemical Energy Conversion and Storage Systems Group, Institute for Power Electronics and Electrical Drives (ISEA), RWTH Aachen University, Germany} \thanks{Jülich Aachen Research Alliance, JARA-Energy, Germany} \thanks{Corresponding author} \and Samuel~Greenbank\footnotemark[1] \thanks{Battery Intelligence Lab, Department of Engineering, University of Oxford, UK}\and Felix~Hildenbrand\footnotemark[2] \footnotemark[3] \and Saad~Jbabdi\thanks{Wellcome Centre for Integrative Neuroimaging, FMRIB, University of Oxford, UK} \and Dirk~Uwe~Sauer\footnotemark[2] \footnotemark[4] \thanks{Helmholtz Institute Münster (HI MS), IEK-12, Forschungszentrum Jülich, Münster, Germany} \and David~A.~Howey\footnotemark[5] \footnotemark[4] }

\maketitle

\section{Abstract} 
Ageing of lithium-ion batteries results in irreversible reduction in performance. Intrinsic variability between cells, caused by manufacturing differences, occurs throughout life and increases with age. Researchers need to know the minimum number of cells they should test to give an accurate representation of population variability, since testing many cells is expensive. In this paper, empirical capacity versus time ageing models were fitted to various degradation datasets for commercially available cells assuming the model parameters could be drawn from a larger population distribution. Using a hierarchical Bayesian approach, we estimated the number of cells required to be tested. Depending on the complexity, ageing models with 1, 2 or 3 parameters respectively required data from at least 9, 11 or 13 cells for a consistent fit. This implies researchers will need to test at least these numbers of cells at each test point in their experiment to capture manufacturing variability.


\section{Introduction}

\label{sec:introduction}
Lithium-ion batteries have grown in importance over the past decade and are now the key technology underlying applications from electric vehicles to grid energy storage \cite{batteryNobel}. High specific energy, low internal resistance and long lifetime have already led Li-ion cells to dominate the market for consumer electronics applications. A crucial issue that strongly impacts overall system performance is the intrinsic variability in capacity, resistance and degradation rate between cells, caused by small variations in manufacturing processes \cite{harris_failure_2017,baumhofer_production_2014,schindler_evolution_2021}. Quantifying the typical variability in mature commercially available Li-ion cells is crucial for understanding the trade-offs involved in designing battery packs and estimating pack performance and lifetime on the basis of individual cell performance.

Battery state of health (SOH) is usually defined as the capacity or impedance/resistance of a cell \cite{rumpf_experimental_2017} under standard test conditions, and it constrains the useful and safe operation of batteries. State of health changes with time and usage, and is influenced both by external factors (such as voltage, time and temperature) and by internal manufacturing and materials variations, \cite{li_data-driven_2019}. Since it has a direct bearing on the economic value and operation of batteries, the estimation of current and future Li-ion SOH is a popular area of research \cite{li_data-driven_2019}.

Variability in cell capacity and resistance can create differing loads within a Li-ion battery pack, and depending on its extent, variability will inevitably impact performance, cost and safety \cite{santhanagopalan_quantifying_2012,dahmardeh_state--charge_2019,orcioni_effects_2017}. There are a variety of sources for this variability, for instance, thermal inhomogeneities influence SOH and increase cell-to-cell variability during usage \cite{rumpf_experimental_2017,liu_effect_2019,paul_analysis_2013}. However, manufacturing variability, sometimes described as tolerance, is a significant contributor to cell-to-cell variability \cite{harris_failure_2017,baumhofer_production_2014,shin_statistical_2015}. New, nominally identical cells from the same batch exhibit a spread in capacity before they have been cycled \cite{schindler_evolution_2021,rothgang_diversion_2014,baumann_parameter_2018}. Simulations and experiments have demonstrated that this intrinsic manufacturing variability is a contributing factor to differing ageing rates between cells \cite{harris_failure_2017,baumhofer_production_2014,rothgang_diversion_2014,kenney_modelling_2012,dubarry_battery_2019}.

There are many possible sources of manufacturing variability \cite{Beck2021}, such as variability in electrode thickness and density \cite{Lenze2018}, fraction of active material, liquid-to-solid ratio and coating gap \cite{Duquesnoy2021}. In this work we assume that all intrinsic variability may be expressed through a single lumped population distribution, for each parameter and dataset. A further contributor to cell-to-cell variability results from variance in experimental conditions, for example location in a given testing chamber \cite{Elliott2020}.

Ageing reduces the performance of a given cell or pack and increases variability between cells \cite{rothgang_diversion_2014,baumann_parameter_2018,liu_effect_2019,schuster_lithium-ion_2015,zilberman_cell--cell_2019,devie_intrinsic_2018}. To compound this, no significant correlation between initial health and ageing rates and subsequent ageing rates in later life has been found experimentally \cite{harris_failure_2017,rothgang_diversion_2014,lucu_data-driven_2020,devie_intrinsic_2018}, although features derived from early cycle life can be used to accurately predict lifetime \cite{severson_data-driven_2019,fermin-cueto_identification_2020}. Some authors have found that initial battery health is sometimes bi-modal  \cite{lehner_reliability_2017,an_rate_2016}. It has also been found that Weibull distributions, common in failure analysis, may be used to quantify battery end of life statistics \cite{harris_failure_2017,schuster_lithium-ion_2015,ossai_statistical_2017,jiang_recognition_2017}. 

In summary, cell-to-cell variability is a significant factor influencing the performance and value of batteries. As a result, modelling of battery ageing data is complicated by the question of how many cells should be tested at each experimental condition so as to adequately capture the intrinsic variability. Battery testing involves cost, in time and number of test channels, and therefore optimising the amount of information obtained from a test is a key consideration. The literature seems to have largely ignored the issue of how many cells should be tested to capture intrinsic variability, and for practical reasons, most ageing studies use only a small number of cells (e.g.\ 1-3) at each test condition. In this paper we address this question directly by fitting models to ageing data. There are many options for modelling battery capacity through life, from empirical curve fits \cite{Ecker2012} through to physical models \cite{reniers_review_2019} and machine learning approaches \cite{li_data-driven_2019,hu_battery_2020}. Since our aim was to investigate intrinsic rather than extrinsic variability, we chose as our modelling approach to use simple empirical curve fits of health versus time. We examined the consistency of the resulting model parameters as we added data drawn from increasing numbers of cells within each dataset, using five different battery ageing datasets. An estimate of required sample size was drawn for each parameter in every model for every available dataset.

\section{Datasets and models}
\label{sec:data_model}
\subsection{Data}
To study cell-to-cell variability, ideally we need data from a very large number of cells, perhaps thousands. The costs of such large scale testing would be prohibitive, requiring many battery test channels for multiple years, and no such datasets are openly available. As a compromise, however, some ageing datasets are available with order 10-100 cells cycled identically (or very similarly). We selected five datasets for analysis based on the requirement of wanting as many cells as possible to have been tested within each dataset. Two of these are open source, and three are from our own experiments. Each individual dataset used identical commercially available Li-ion cells, albeit having different manufacturers, chemistries and cell sizes from dataset to dataset. All datasets used 18650 cylindrical cells, although the methods discussed below can equally be applied to other form factors. Some of the datasets featured identical experimental conditions, i.e.\ each cell was tested in exactly the same way, whereas others varied the testing conditions slightly beyond the expected uncontrollable experimental variability. The datasets are as follows:

\begin{enumerate}
	\item \textbf{Baumh\"ofer-2014} \cite{baumhofer_production_2014} consists of 48 Sanyo/Panasonic UR18650E NMC/graphite 1.85 Ah cells in a cycle ageing test each under the same operating conditions. Data available at \cite{DataBaumhoefer2014}.
	\item \textbf{Dechent-2020} consists of 22 Samsung INR18650-35E NCA/graphite cells each with a nominal capacity of 3.4 Ah. The cells were cycled with C/2 constant current and a 20\% cycle depth around an average SOC of 50\%.  Data available at \cite{Dechent2020}.
	\item \textbf{Dechent-2017} consists of 21 Samsung NR18650-15L1 1.5 Ah NMC/graphite cells. Six of the cells were cycled with 1C charge and 6C discharge current between 3.3 V and 4.1 V (90\% cycle depth), and 15 additional cells were cycled with the same voltage range but current rates varied by up to 15\%.  Data available at \cite{Dechent2017}.
	\item \textbf{Severson-2019} consists of 124 cells made by A123 APR18650M1A with LFP/graphite chemistry, each with nominal capacity 1.1 Ah. For this work a subset 67 of these cells with similar load profiles but slightly varying charging currents was chosen. Data available via \cite{severson_data-driven_2019}. The cells in this dataset are from three different experimental batches so will have been subjected to higher variance in testing conditions than the other datasets.
	\item \textbf{Attia-2020} replicates Severson-2019, but with a fixed charging window of 10 minutes. There are 45 cells in this set. Data available via \cite{attia_closed-loop_2020}.
\end{enumerate}

Three of these datasets, namely Baumh\"ofer-2014, Severson-2019 and Attia-2020, exhibit an onset of rapid degradation in later life, sometimes called the `knee-point' \cite{baumhofer_production_2014,fermin-cueto_identification_2020}. The other two datasets show only linear degradation over usage. The data prior to the knee-point in the Attia-2020 and Severson-2019 sets was separately extracted to produce two additional linear ageing datasets.

\subsection{Models}

The models and the corresponding datasets that they were fitted to are shown in Table \ref{tab:dataset_model_combinations}. Here, Linear-1 and Linear-2 refer to the two linear models, having one and two parameters respectively. Alternatively, LinExp is a combined linear and exponential model that was used to capture the knee-point and later life health decay, where this was evident in the data.

\begin{table}[]
    \centering
    \begin{tabular}{|c|c|c|c|c|}
    \hline
    Dataset & Linear-1 & Linear-2 & LinExp & Ref.\ \\ \hline
    Dechent-2017 & X & X & & \cite{Dechent2017} \\
    Dechent-2020 & X & X & & \cite{Dechent2020} \\
    Baumh\"ofer-2014 & & & X & \cite{baumhofer_production_2014} \\
    Severson-2019 & X & X & X & \cite{severson_data-driven_2019} \\
    Attia-2020 & X & X & X & \cite{attia_closed-loop_2020} \\ \hline
\end{tabular}
    \caption{The dataset and empirical model combinations used here. Linear-1 is a single parameter linear model, Linear-2 a two parameter linear model, and LinExp a 3 parameter linear plus exponential model.}
    \label{tab:dataset_model_combinations}
\end{table}

\begin{figure}
    \centering
    \includegraphics[width=\textwidth]{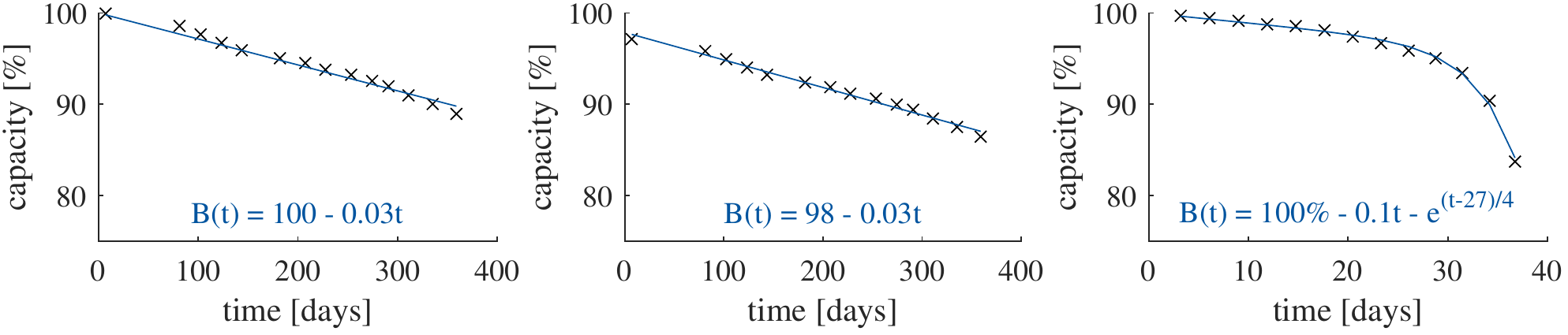}
    \caption{Examples of data and fitted curves. Left to right: Linear-1 model on a cell from Dechent-2020; Linear-2 model on a cell from Dechent-2020; LinExp model on a cell from Severson-2019.}
    \label{fig:example_model_fitting}
\end{figure}

The three models are given by the following expressions, where $t$ is time, $B$ is capacity, and all other parameters $(c_1, c_2, c_3, B_0, t_f, \tau)$ were fitted to the data:
\begin{align}\label{eqn:linearmodelone} 
	&\text{Linear-1:}& B(t)&=100\% + c_1 \times t \\
    &\text{Linear-2:}& B(t)&=B_0 + c_2 \times t \\
	&\text{LinExp:}  & B(t)&=100\% + c_3 \times t - \exp{\left[\left(t-t_f \right) /\tau \right]} 
\end{align}

Linear-1 and Linear-2 differ only by the addition of the initial capacity $B_0$ as a fitted parameter in the latter. The cell capacities were normalised according to which model was in use. For Linear-1 and LinExp, the capacities were normalised relative to the initial capacity of each cell. Linear-2 used capacity curves normalised relative to the nominal capacity. In the LinExp model, the initial linear capacity decrease is followed by a faster exponential decrease with onset time $t_f$ and time constant $\tau$, as shown in Fig.\ \ref{fig:example_model_fitting}.

\section{Methodology}\label{sec:approaches}

\begin{figure}
\centering
    \begin{subfigure}[b]{0.2\textwidth} \centering
	\includegraphics[width=.8\textwidth]{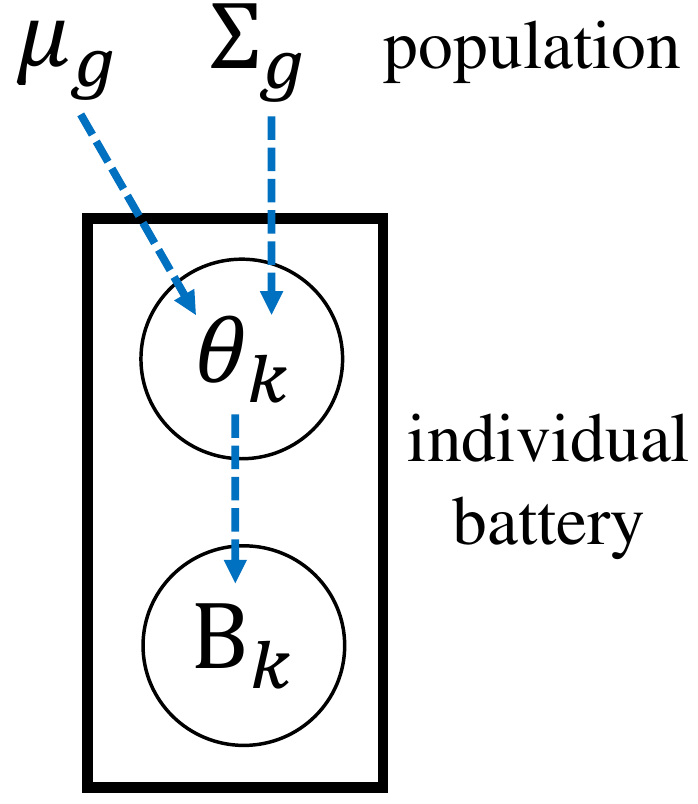}
	\caption{Hierarchical model}
	\label{fig:MLB_diagram}   
    \end{subfigure}     
    \hfill
    \begin{subfigure}[b]{0.45\textwidth}
	\includegraphics[width=.95\textwidth]{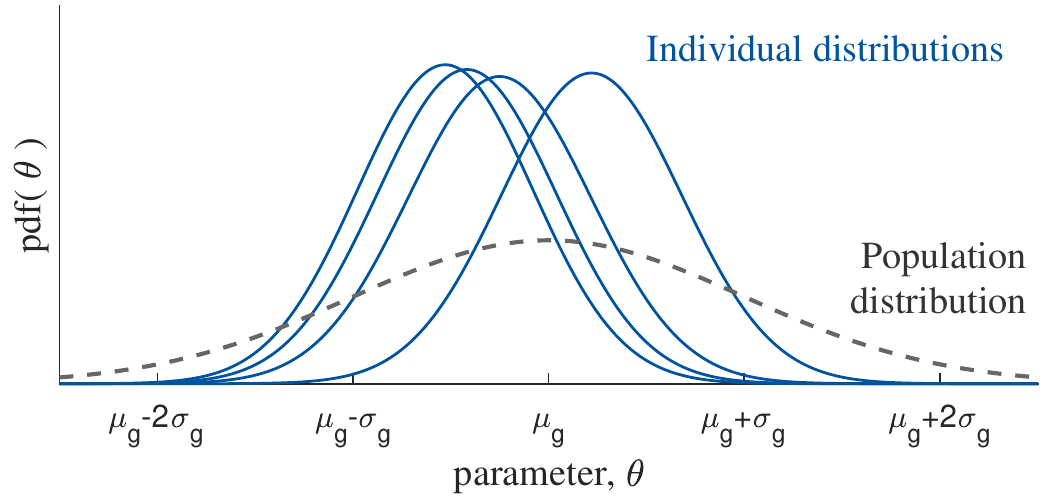}
	\caption{The relationship between population distribution and distributions of individual cells/samples.}
	\label{fig:MLB_global_individual_plot}
    \end{subfigure}
    \hfill
    \begin{subfigure}[b]{0.3\textwidth}
	\includegraphics[width=.90\textwidth]{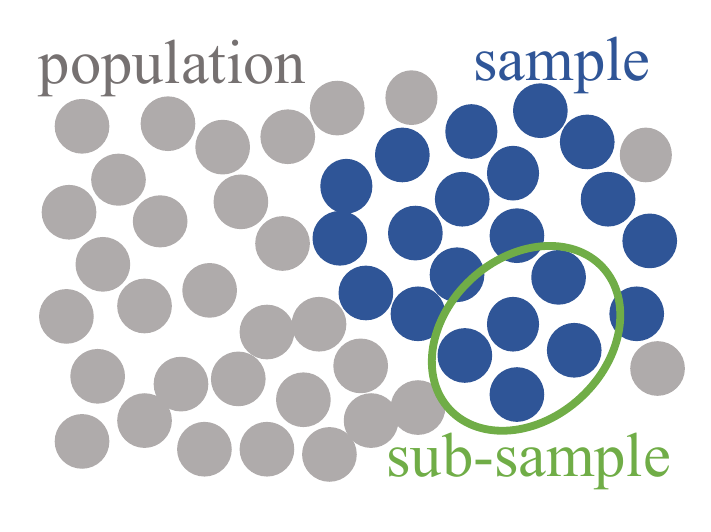}
	\caption{A sub-sample is part of a sample, which is part of a population.}
	\label{fig:MLB_global_individual_bubbles_plot}
    \end{subfigure}
\caption{A hierarchical approach was used to infer population statistics.}
\label{fig:MLB_diagram_and_plots}
\end{figure}

\begin{figure}
    \centering
    \includegraphics[width=.6\textwidth]{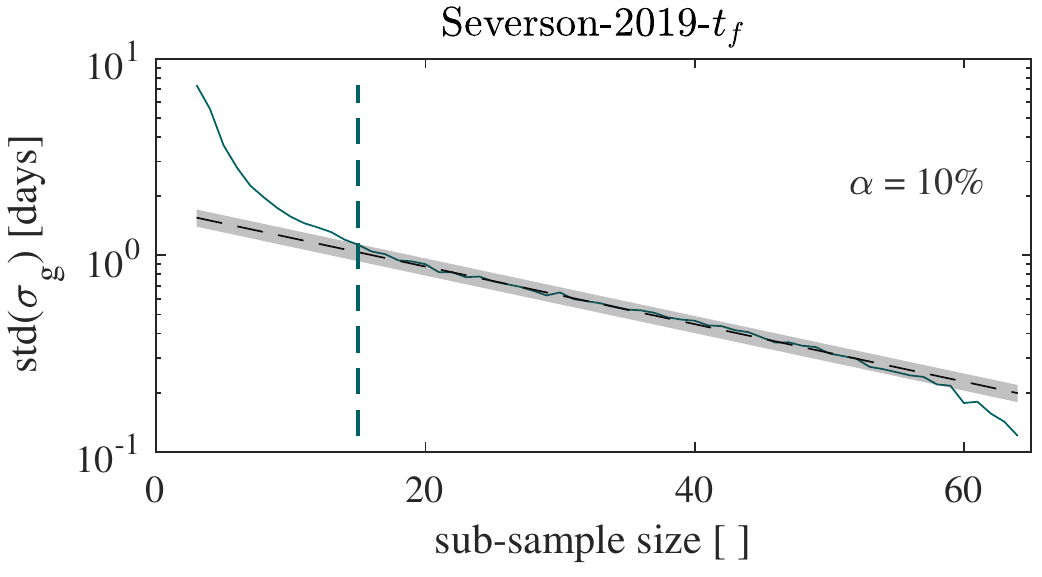}
    \caption{Example showing the decreasing standard deviation of the population standard deviation estimate as the number of sub-samples used for model fitting is increased. A threshold for an acceptable estimate is shown (vertical dashed line). Example taken from the $t_f$ parameter for Severson-2019.}
    \label{fig:cellsrequired_calculation}
\end{figure}

To quantify cell-to-cell variability, we used an approach called multi-level Bayes (MLB), also known as hierarchical Bayes, where the parameters of an individual cell model are assumed to be drawn from a population distribution, as depicted in Fig.\ \ref{fig:MLB_diagram}. In this framework, the first level of inference is on the parameters of an individual battery cell model, and the second level of inference is on the parameters of the underlying population distribution \cite{Woolrich2004,murphy_machine_2012,theodoridis_chapter_2020}. Given some data sub-sampled from the datasets described above, this approach provides an estimate of the individual ($\theta_k$) and the population ($\mu_g, \Sigma_g$) parameter values as well as their associated uncertainties, as depicted in Fig.\ \ref{fig:MLB_diagram_and_plots}(a) and (b). Therefore one can explore the trade off between the number of cells' data used for fitting the models versus the stability and variance (or standard deviation) of the resulting population parameter estimates. As additional data from more cells is included in the estimation, the variance of the population mean and variance decreases (i.e.\ we become more certain of the population model). As illustrated in Fig.\ \ref{fig:cellsrequired_calculation}, we considered the population estimates to be stable when the standard deviation of the population standard deviation estimate began to steadily decrease as a function of sub-sample size ($\sim \frac{1}{N}$). We set the condition of an acceptable variability as being within a threshold, $\alpha$, of the stable decreasing region. The value of $\alpha$ was set at $10\%$ as shown by the grey shaded region in Fig.\ \ref{fig:cellsrequired_calculation}. 

The following conventions are used throughout the remainder of this work. Fig.\ \ref{fig:MLB_diagram_and_plots}(c) shows the definitions of population, sample and subsample used. The `population' means the very large (but unavailable) group of all possible similar batteries produced in the same manufacturing batch, from which a subset were tested in a lab. (Therefore, we expect the population statistics to be different for each dataset that was introduced in the previous section.) A `sample' refers to all the available full data in a specific dataset. Therefore, a sample is drawn from a population. Conversely, any time a smaller subset was drawn from a full test dataset, it is referred to here as a `sub-sample'. Summary sample statistics are denoted with the Latin alphabet, while population estimates are written using the Greek alphabet. For example, mean and variance are $(m,s^2)$ and $(\mu,\sigma^2)$ respectively. The letter $k$ is used to denote value(s) for a specific cell, and $K$ denotes the total number of cells. Probabilities and distributions are written in capitals: $\mathrm{P}$, $\mathrm{N}$. Battery capacities are represented by the letter $B$. 

\subsection{Multi-Level Bayes}\label{sec:multi-level-bayes}

The parameters of an individual cell degradation model are assumed to be drawn from a population distribution that is unique for each dataset, with unknown population mean and variance. Let $B_k$ denote the capacity of cell $k$ over a number of measurements (i.e.\ $B_k$ is a vector) and $\theta_k$ denote the model parameters determining the time evolution of capacity. For example, for the LinExp model, $\theta_k = (c_k,t_{f,k},\tau_k)^T$. Assuming additive and Gaussian measurement noise, 

\begin{equation}
    B_k = f(\theta_k) + \epsilon_k,
\end{equation}
where $\epsilon_k \sim \mathrm{N(0,\sigma_{n,k}^2)}$ and $f$ is one of the three models introduced above.

The parameters of an individual cell model $k$ are themselves drawn from a population distribution, assumed here to be Gaussian,

\begin{equation}
    \theta_k \sim \mathrm{N(\mu_g,\sigma_g^2)},
\end{equation}
where $\mu_g$ is the group (i.e.\ population) mean and $\sigma_g^2$ the group variance. These population parameters are vectors of one, two, or three elements depending on whether the generative model is Linear-1, Linear-2, or LinExp, respectively.

To complete the specification of this generative model in the Bayesian framework requires prior distributions to be assumed. We used wide Gaussian priors (zero mean, $10^4$ variance) on the population means $\mu_g$, and uniform distributions on the population variances. The noise parameters $\sigma_{n,k}$ were assumed to follow a Jeffrey's prior, $\mathrm{P(\sigma_{n,k}^2)}\sim1/\sigma_{n,k}^2$, and were integrated out analytically.

To infer the posterior distributions of the population parameters $\mu_g$ and $\sigma_g$, we used a two step process. In the first step (first level inference), we fitted individual cell parameters $\theta_k$ using Markov chain Monte Carlo (MCMC) sampling to obtain samples from $\mathrm{P(B_k|\theta_k)}=\int \mathrm{P(B_k|\theta_k,\sigma_{n,k})}\mathrm{P(\sigma_{n,k})}d\sigma_{n,k}$. We then approximated the distributions for each cell with a Gaussian using their summary statistics (means $\mu_k$ and variances $\sigma_k^2$), which were then used in a second step (second level inference). This allowed the full posterior distribution of the population parameters to be written as follows:

\begin{equation}
    \mathrm{P(\mu_g,\sigma_g | \{B_k\})} = \int \dots \int \mathrm{P(B_k|\theta_k)}\mathrm{P(\theta_k|\mu_g,\sigma_g)}\mathrm{P(\mu_g,\sigma_g)}d\theta_1...d\theta_{K}
\end{equation}

The above multivariate integral can be evaluated analytically, owing to the Gaussian approximation to the first level inference. This yields:

\begin{equation}
    \mathrm{P(\mu_g,\sigma_g | \{B_k\})} \propto \prod_{k=1}^{K} \mathrm{N(\mu_k,\sigma_k^2+\sigma_g^2)} \mathrm{P(\mu_g,\sigma_g)}
\end{equation}

Intuitively, this last expression shows that the posterior population mean is a Gaussian centred around the weighted average of the individual cell parameters, where the weights combine the first level variances (uncertainty on the fitting of each cell) and the group variance (mixed effect model). After this, we again used MCMC to draw samples from this posterior distribution to calculate its summary statistics.

The MLB approach was used to fit the parameters of all the model/dataset combinations shown in Table \ref{tab:dataset_model_combinations}. That fitting was performed at all sub-sample sizes from minimum 3 cells, up to 3 less than the full number of cells in each dataset, with 1,000 repeats performed at each sub-sample size using random selection with replacement. A population distribution was deemed to have a stable fit if the standard deviation of the estimate of $\sigma_g$ settled to follow a function $\log y = a x + b$ where $x$ is sub-sample size, $y$ is the standard deviation of $\sigma_g$ and $a,b$ are arbitrary slope and offset parameters. As a comparison, the equivalent result was also plotted when using a simple sub-sample distribution (SSD) by taking the mean, $m_g$, and variance, $s_g^2$, of a given sub-sample:

\begin{align}
    m_g = \frac{1}{K} \sum_{k=1}^{K} \theta_k, \quad s_g = \sqrt{ \frac{\sum_{k=1}^{K}\left(\theta_k - m_g\right)^2 }{K-1} }
\end{align}

\section{Results}
As a reminder, the objective is to quantify the number of battery cells that are required for a stable fit of a population model, when cells are selected at random from a population. In particular, we wish to infer both the parameters of the capacity fade model for each cell, and the parameters of the underlying population, including their uncertainties. We now examine both aspects in succession across the various datasets and models.

\begin{figure}
    \centering
    \includegraphics[width=\textwidth]{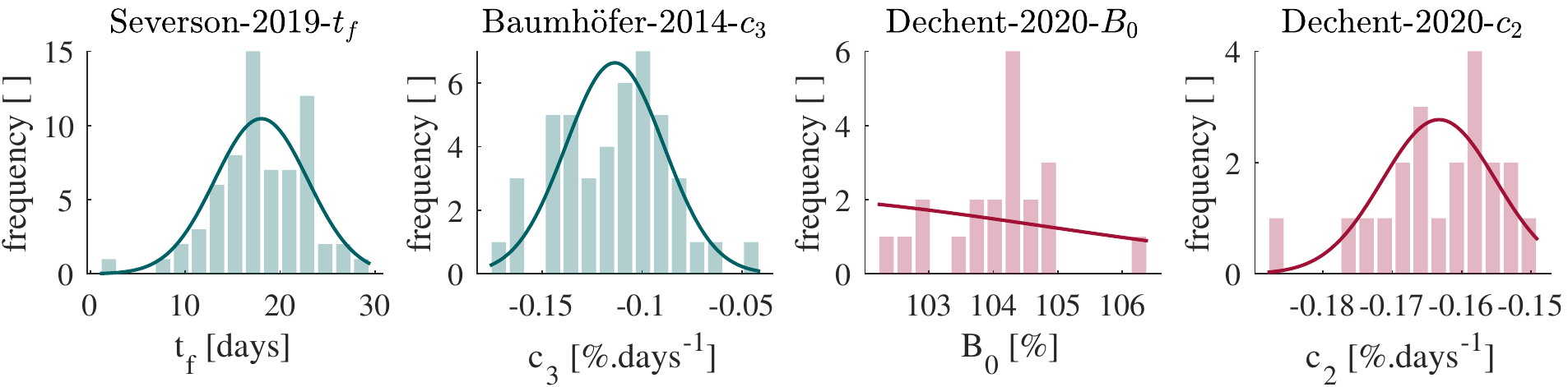}
    \caption{Generally, the inferred population distributions (lines) and sample distributions (bars) are similar, but there are some exceptions.}
    \label{fig:sample_versus_population_distribution}
\end{figure}

In most cases, the parameter values estimated by MLB for the population level distributions match that of the sample distributions. Fig.\ \ref{fig:sample_versus_population_distribution} shows two examples of well matched population and sample distributions, namely from Severson-2019-$t_f$ and Baumh\"ofer-2014-$c_3$. The parameter distribution in the sample is approximately Gaussian in both cases, matching an important assumption in the MLB derivation. 

On the other hand, Dechent-2020 with a Linear-2 model demonstrates how the population level parameters can adjust to accommodate non-Gaussian distributions. In the case of Dechent-2020-$B_0$, this manifests as a wide distribution over $B_0$ and $\mu_g$ offset from the sample. The rest of the data was fit more precisely, resulting in a more representative distribution over the gradient parameter $c_2$.

There was very little correlation between parameter distributions for any dataset/model combinations. The only significant case was for the Attia-2020-LinExp model, where Pearson's rank coefficients of 0.94 were found between all of $c_1$, $t_f$ and $\tau$. Severson-2019 had correlation coefficients of $\sim$0.6, suggestive of a weak relationship between the parameters of the LinExp model.

\begin{figure}
    \centering
    \includegraphics[width=\textwidth]{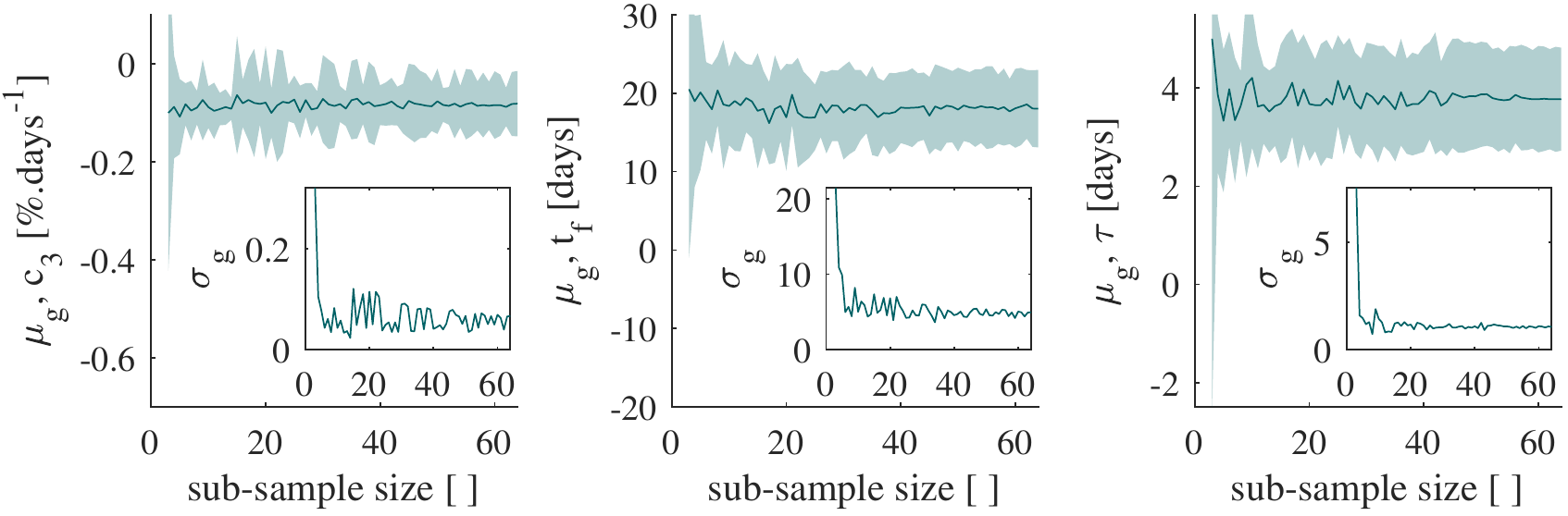}
    \caption{MLB results when fitting a single sub-sample draw have significant variability, shown here using Severson-2019 with the LinExp model. Lines show mean parameter estimates, shaded region is 1-$\sigma$  about that mean, inset is $\sigma_g$.}
    \label{fig:single_shot_fitting_by_subset_size}
\end{figure}

The MLB results were subject to significant variability when fitting a single incidence at each sub-sample size. Fig.\ \ref{fig:single_shot_fitting_by_subset_size} demonstrates a typical set of parameter estimates where it was hard to interpret the values at small sub-sample sizes. The mean estimates appeared to be well fit at small values, but the population distributions only settled when the majority of the sample was used.

\begin{figure}
    \centering
    \includegraphics[width=\textwidth]{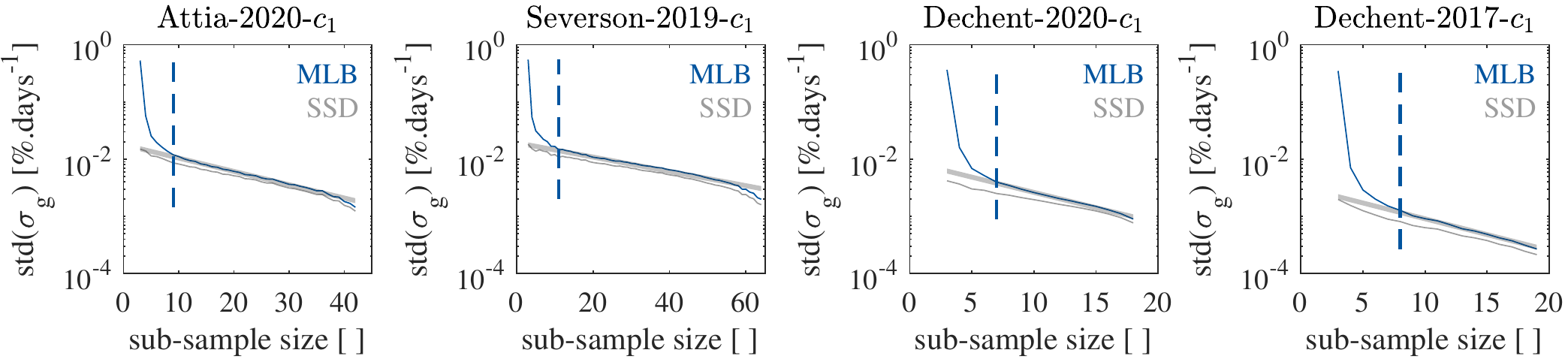}
    \caption{Smooth standard deviations are estimated by taking 1,000 random subsamples (with replacement), shown here with the Linear-1 model. MLB is Multi-Level-Bayes, SSD is sub-sample distribution, over the number of cells in the sub-sample.}
    \label{fig:results_Linear-1}
\end{figure}

The summary results from 1,000 repeats, with replacement, were much smoother. The estimated standard deviation of $\sigma_g$ for the Linear-1 models rapidly dropped with increasing numbers of cells in a sub-sample for all datasets as shown in Fig.\ \ref{fig:results_Linear-1}. The SSD approach produced a lower variance at all sub-sample sizes, but appears insensitive to small sub-samples.

\begin{figure}
    \centering
    \includegraphics[width=\textwidth]{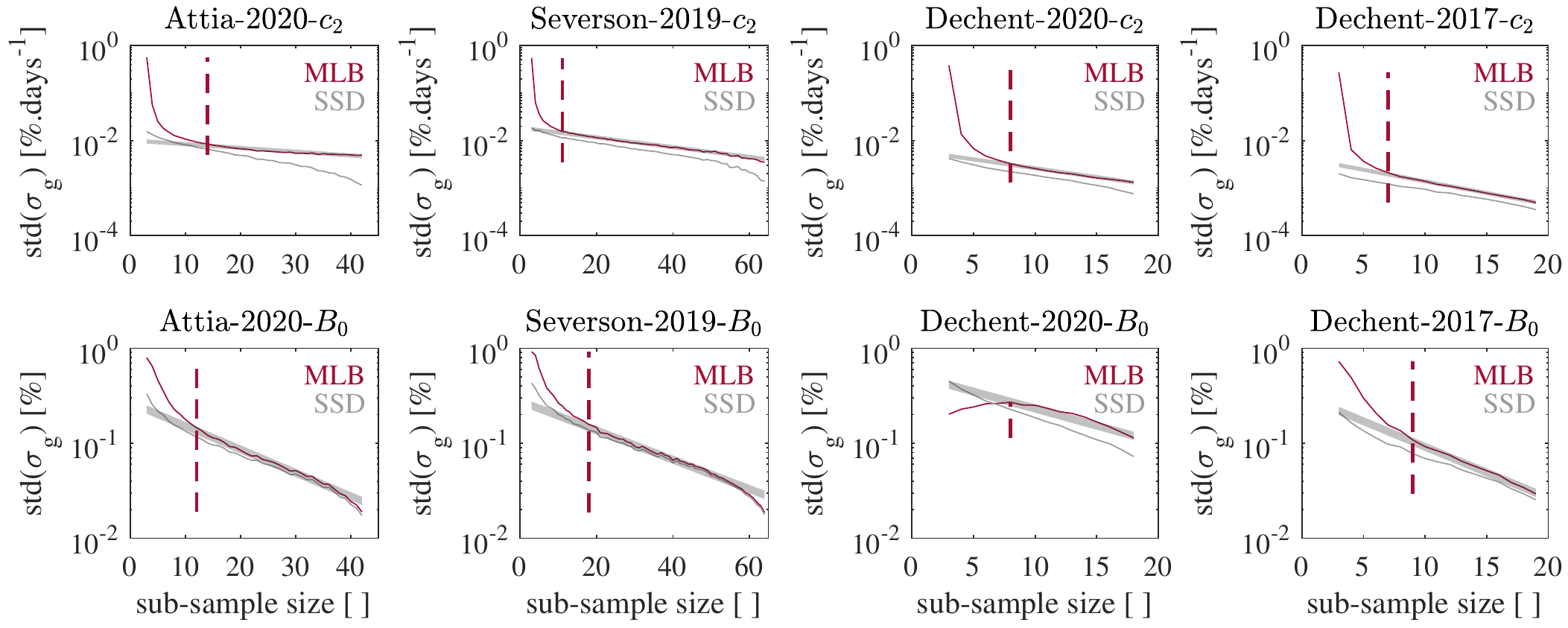}
    \caption{The standard deviations of the population-level standard deviation estimates for the Linear-2 model. MLB shows the Multi-Level-Bayes and SSD the sub-sample distribution results over the number of cells in the sub-sample.}
    \label{fig:results_Linear-2}
\end{figure}

\begin{figure}
    \centering
    \includegraphics[width=\textwidth]{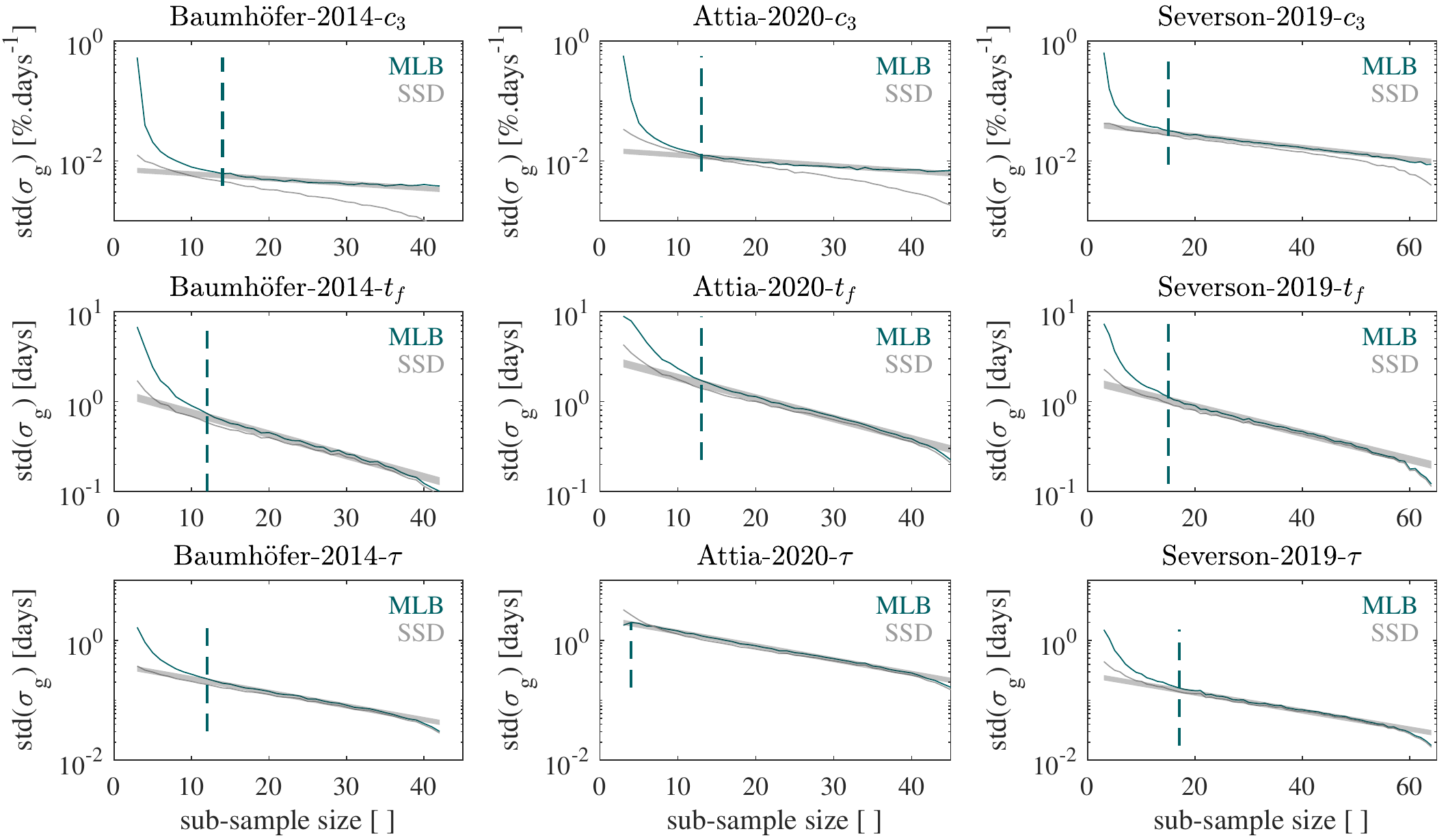}
    \caption{The standard deviations of the population-level standard deviation estimates for the LinExp model. MLB shows the Multi-Level-Bayes and SSD the sub-sample distribution results over the number of cells in the sub-sample.}
    \label{fig:results_LinExp}
\end{figure}

The results for Linear-2 and LinExp were very similar as shown in Figs.\ \ref{fig:results_Linear-2} and \ref{fig:results_LinExp}, although there were distinctly less stable fits for Dechent-2020-$B_0$ and Attia-2020-$\tau$. All three models shared a reduced standard deviation of $\sigma_g$ when using SSD.

The linear relationship between sub-sample size and the log of the standard deviations was deemed to represent a consistent fit. It was subsequently used to determine when an `effective' sub-sample size had been reached. A model was considered well fit when the standard deviation of $\sigma_g$ was within $\alpha=10\%$ of this stable section, found using a linear extrapolation (as plotted in the figures).

\begin{figure}
    \centering
    \includegraphics[width=.6\textwidth]{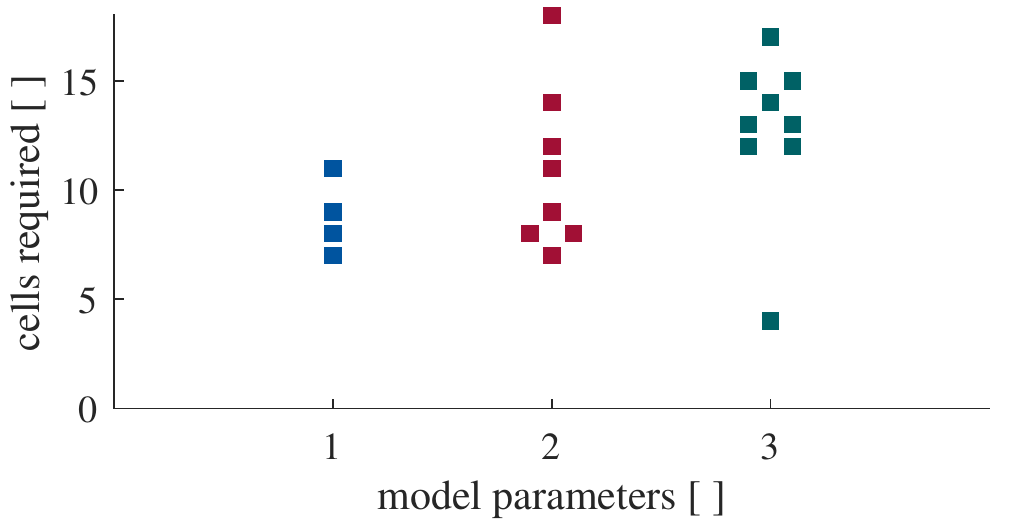}
    \caption{Testing with order 10 cells is required to achieve stable population estimates, with the number increasing as the model complexity increases. (Offset x-axis values were used to show identical results.)}
    \label{fig:cellsrequired}
\end{figure}

Fig.\ \ref{fig:cellsrequired} shows the relationship between the number of cells required to achieve `stable' population estimates, vs.\ the number of model parameters. The number is shown for all model, dataset and parameter combinations. The mean required sub-sample sizes for a consistent fit were 9, 11 and 13 for the 1, 2 and 3 parameter models respectively.

\section{Discussion}

The number of cells required to fit the various models presented here and capture a stable estimate of the population variability is of order 10. For the simplest model, Linear-1, the number was 9 cells, and for the most complex LinExp model, the number increases to 13. The results understandably suggest that increased model complexity leads to an increase in the number of cells required to be tested to achieve a stable estimate of the population variability.

The multi-level Bayesian approach produced consistent parameter estimates from sub-samples. Given that cell-to-cell variability is an important phenomenon impacting battery performance, the estimated distributions are an invaluable tool to use in empirical modelling. Simple sample distribution techniques are limited to estimates of spread within the domain of the sample and hence showed less sensitivity to sub-sample size here when using random selection. The number of cells required to estimate population variability was fairly consistent across the datasets investigated here and was a stronger function of the model complexity than of the dataset. However, future work could test the robustness of this conclusion across a wider range of datasets.

The standard deviation of the $\sigma_g$ estimates reduced as sub-sample sizes were increased. In most cases, the SSD and MLB results also approached the same values as sub-sample sizes increased because the two techniques will return similar results at high sub-sample sizes. At low sub-sample sizes SSD was limited to the variability of the sub-sample, whereas MLB was less certain, resulting in higher values for both $\sigma_g$ and its standard deviation. In this case, SSD appears to have been artificially confident as an estimate of the population distribution.

The chosen threshold condition for a well fit $\sigma_g$ parameter resulted in consistent results. The same consistency was also found when using other threshold values of $\alpha$. The hypothesis that sub-sample size increases with model complexity appears to be supported. However, it would be useful to explore this in more depth using larger datasets.

In the derivation of the MLB approach, we assumed there to be no correlations between parameters in the prior probability distributions. That assumption was found to be questionable in two cases here. Future work should explore the impact of this on population modelling.

The results for the Dechent-2020 dataset with the Linear-2 model demonstrated the robustness of the MLB approach by fitting a very similar gradient to the Linear-1 version, despite an apparently uncertain value of $B_0$. The estimate of $\mu_g$ for $B_0$ was 99.7\%, i.e.\ the resultant population model was very similar to that with Linear-1. Our current approach assumes a Gaussian distribution at the population level. In the case of a bi-modal (or multi-modal) population distribution, we expect that the MLB method would respond by estimating a wide standard deviation, but this has not been tested. Extending the method proposed in this paper to other population distributions would be an interesting subject for future study. 

Various ageing mechanisms are likely responsible for the degradation datasets considered in this paper. In the case of the Severson dataset, it is likely that degradation was largely caused by lithium plating \cite{severson_data-driven_2019}, while on the Dechent-2020 dataset, covering layer formation and jellyroll deformation are key to degradation \cite{Willenberg2020}.

Between the datasets small differences were found. For the Linear-2 model the datasets Attia-2020 and Severson-2019 show a higher cell number threshold required for population estimation. The reasons for the trends we see are varied. The underlying mechanisms may be caused by increased variation of cycling conditions within the dataset. In order to capture higher usage variation in addition to intrinsic variability the number of cells will likely need to be increased.  Future research could look into quantifying both use variability and manufacturing variability at the same time.

The fact that more complex models required more cells to be tested at each test point is challenging for battery lifetime experiments, since it could increase greatly the number of test channels and cells required in long term ageing experiments. We did not extrapolate to higher numbers of parameters or to other models, but it is reasonable to assume that the issues explored here will be present in other, more complex cases. 

One challenge with the technique used here is that it relied on limited size samples from the population. Future work could explore whether larger sample sizes lead to similar results as found here.

Finally, further work is required to investigate the impacts of cell-to-cell variability using more complex physics-based models of battery ageing \cite{reniers_review_2019}, which can have 5-10 or more parameters in addition to the 20+ parameters of the required underlying electrochemical model. Openly available ageing datasets are at present too small to enable meaningful calculation of population parameter variability for such complex models using the methods we outline, since the number of parameters is significantly more than 1-3. One approach in a future study could be using synthetic datas to study more complex models with more parameters and test parameter identifiability \cite{Aitio2020}.  

\section{Conclusions}

Simple empirical battery capacity fade models were fitted to a variety of ageing datasets to quantify the number of cells required to estimate the variability of the underlying population. The number of cells required to give a stable population variance estimate was found to vary according to the number of parameters in a given model. Respectively, 9, 11 and 13 cells are estimated to be required for models with 1, 2 and 3 parameters. Both sample statistics and population estimates were shown to stabilise with under 20 cells in most cases but this relied on there existing a Gaussian distribution of parameters within the sample, otherwise 20 cells were required.

For capacity curve fitting, perhaps the biggest challenge going forward is the selection of appropriate ageing model order and structure. This should be done not just by looking to what functions fit the capacity profiles best, but which functions produce the most reliable parameter distributions when looking at a dataset as a whole.

There was insufficient data here to test these results and conclusions as a function of variability caused by differences in usage, but this would be an interesting future exploration topic. Also, model selection across larger datasets is a challenging problem. For example, some of the battery capacity fade trajectories in this study fitted well to a linear degradation stage followed by an exponential decay starting from some knee point. However some of the resultant sample distributions cannot be confidently used to calculate basic summary statistics, such as Dechent-2020-$B_0$.

Understanding and quantifying battery cell-to-cell manufacturing variability is an open research topic, and this work represents an initial step. These results form a useful order of magnitude guide, for those undertaking long term battery ageing experiments, of what is needed to capture manufacturing variability.


\section*{Acknowledgment}

The authors would like to thank P.M. Attia, K. Severson and coworkers \cite{severson_data-driven_2019} and \cite{attia_closed-loop_2020}. S.G. is funded by EPSRC, UK, and Siemens Ltd. F.H gratefully acknowledges the financial support of Engie. P.D., F.H. and D.U.S gratefully acknowledge the financial support by Bundesministerium für Bildung und Forschung (BMBF 03XP0302C, 03XP0320A) and Engie.
\section*{Orcid}
Philipp Dechent: 0000-0003-3041-1436\newline
Samuel Greenbank: 0000-0002-2091-717X\newline
Felix Hildenbrand: 0000-0002-7819-3223\newline
Saad Jbabdi: 0000-0003-3234-5639\newline
Dirk Uwe Sauer: 0000-0002-5622-3591\newline
David Howey: 0000-0002-0620-3955
\section*{Author contributions}

P.D. and S.G.: conceptualisation, software, data curation, writing – original draft, visualisation.
F.H.: software, writing – review and editing.
S.J.: methodology, software, writing – review and editing.
D.U.S.: supervision, funding acquisition.
D.A.H.: conceptualisation, supervision, funding acquisition, writing – review and editing.

\section*{Conflict of Interest}
The authors declare no conflict of interest.

\section*{Keywords}

battery, ageing, lithium-ion, degradation, statistics, manufacturing, testing, energy conversion, electrochemistry

\section*{Bibliography}
\bibliographystyle{unsrt}
\bibliography{references.bib}

\end{document}